# Wrapping trust for interoperability. A study of wrapped tokens

Caldarelli Giulio

University of Verona. Department of Business Administration

Via Cantarane 24, 37129, Verona, Italy

giulio.caldarelli@univr.it

**Abstract**

As known, blockchains are traditionally blind to the real world. This implies the reliance on third parties called oracles when extrinsic data is needed for smart contracts. However, reintroducing trust and single point of failure, oracles implementation is still controversial and debated. The blindness to the real world makes blockchains also unable to communicate with each other preventing any form of interoperability. An early approach to the interoperability issue is constituted by "wrapped tokens," representing blockchain native tokens issued on a non-native blockchain. Similar to how oracles reintroduce trust, and single point of failure, the issuance of wrapped tokens involves third parties whose characteristics need to be considered when evaluating the advantages of "crossing-chains". This paper provides an overview of the wrapped tokens and the main technologies implemented in their issuance. Advantages, as well as limitations, are also listed and discussed.

## 1. Introduction

"*Bitcoin can't speak the language of Ethereum and vice versa……we can't spend bitcoin on the Ethereum network, nor can we make use of Ethereum's smart contracts on the Bitcoin network*" [1]. There is no one-size-fits-all solution for blockchains, and each distinct ecosystem offers differences in scalability, security, programmability, and privacy. The different needs of their users then justbiify the heterogeneity of coexisting separate platforms, but from a financial perspective, their inability to communicate means that the related capital is not completely exploitable for DeFi purposes [2]. As a matter of fact, interoperability across chains wasn't integrated into the original idea of blockchain protocols; however, as technology grew, the need for cross-chain solutions arose [3]. Interoperability would grant users accessibility, speed, and fees reduction. Having BTC, the most hyped crypto asset on the Ethereum network, for example, and including it in smart contracts was quite a fascinating idea, and it has been possible thanks to the introduction of wrapped tokens [4][5]. Wrapped tokens were proposed as a means to overcome the absence of communication between different blockchains [6]. According to the DeFi Pulse website, one of the most reliable resources for statistic analysis on DeFI, it emerges that more than 270,000 BTC (September 2021) are actually being used in DeFi as Wrapped Tokens [7]. The same website also shows that 80% of wrapped bitcoins are WBTC, while the remaining part is provided by other or emerging companies/platforms. It is a huge amount, considering how much bitcoin is worth, but since the total circulation of bitcoin is over 18M, those used as wrapped tokens are still less than 2%. Lately, the offer of wrapped tokens is increasing, and many companies such as Tron, Binance, Klever, Solana, and Tezos are offering their cross-chain proposal [6], [8], [9]. Although all of these are based on the implementation of wrapped tokens, the bridging process slightly differs. With the technology available today, bridging an asset to another blockchain means locking the original asset on its own blockchain and minting a copy on another that is not naturally compatible [10].

This paper aims to explain the idea behind the creation of wrapped tokens, outlining their role, use, and drawbacks. The first known examples, such as WETH, WBTC renBTC will be discussed to understand the differences among the entities involved in their management [11]–[13]. Other approaches such as Synths and Layer2 solutions will be discussed, analyzing the case of Synthetix and Secret network [14], [15]. The scope of this paper is to give a broad overview of the concept of wrapped tokens to facilitate further theoretical and empirical studies on this subject. In the absence of academic resources on the subject, only grey literature could be reviewed. From the analyzed examples, it emerges that a real bridge between blockchains still doesen't exist, but wrapped tokens constitute an effective workaround till this feature is developed. The paper proceeds as follows. The next section outlines the history of wrapped tokens and their predecessors. Section three outlines some of the most promising solutions in the field, while section four discusses their characteristics, analyzing strengths and weaknesses. Section five concludes the paper by providing hints for further research.

## 2. Understanding wrapped tokens

To better understand the concept of wrapped tokens, it may be helpful to introduce the concept of wrapped "asset" first. The wrapped asset is a token that represents a real-world or crypto asset, and it is backed by the represented asset or assets of equal value [16]. The backing asset is put in a vault called a "wrapper" (hence wrapped asset) [10]. It is issued on a blockchain, and it is supposed to keep the same value as the represented asset. Tokenizing an asset means converting its structure to a blockchain-compliant one so that smart contracts can manage it. The underlying idea of a wrapped asset is that since it is impossible to physically move an asset to the blockchain or make it blockchain compliant, a token representing the asset is created instead. Of course, to avoid double-spending, the real-world assets should be locked till it's "avatar" is used on the blockchain. Examples of wrapped assets are the so-called stablecoins [17]. The stablecoins are tokens that represent stable assets and are particularly useful in DeFi, given the usual volatility of cryptocurrencies. In blockchain-based financial applications, fiat currencies were considered useful to hedge against the risk of volatility. Unfortunately, those could not be directly imported into smart contracts for being in the real world and having no blockchain-compliant characteristics. Therefore, in order to have those stable assets on the blockchain, a representation of these had to be created with blockchain-compliant characteristics. USD-T and USD-C are two stablecoins, respectively, issued by Tether Foundation and Centre organization [18], [19]. The issuing organizations are in charge of guaranteeing the peg of those cryptocurrencies to the US dollar. In principle, in order to mint stablecoins, an equivalent amount of dollars or financial assets should have been "held" by trusted entities chosen by the issuing companies. The centralization of control of those currencies and the absence of a transparent proof of assets makes communities reluctant to consider those as properly wrapped assets [20]. Stablecoins that are more widely recognized as wrapped assets are synthetic or decentralized ones such as sUSD, DAI, USD-J [21], [22]. The sUSD is a stablecoin issued by Synthetix when staking SNX native token and has a collateralized ratio of 400%. It means that to mint 10 sUSD, at least 40$ of SNX needs to be locked. DAI and USD-J, are decentralized projects issued by MakerDAO and JUST-GOV, respectively. Those allow the issuance of stablecoins by collateralizing assets into a smart contract with at least 150% ratio. The main difference between the first group of stablecoins (USD-T, USD-C) and the second one (sUSD, DAI, USD-J) is that; while the first involves a centralized authority and does not offer a transparent proof of asset, the second is fully managed by smart contracts and the proof of asset is available on-chain.

Wrapped tokens are a particular category of wrapped assets that is not intended to create a crypto representation of real-world assets such as gold, stock, or fiat currencies but of other crypto assets instead (e.g., Bitcoin, BNB, Ether) [23]. The reason for their existence is because, as blockchains are entirely isolated from the external world, they are also isolated from each other. Therefore, as there is no way to bring dollars on Ethereum blockchain and stablecoins are used instead, there is also no means to bring bitcoin to Ethereum, and wrapped tokens must be used instead [24]. Although hypothesized and desirable, the communication and interoperability between blockchain are impossible with the actual technologies, and cryptocurrencies cannot "jump" from one chain to another. The limitation of interoperability is then addressed with the same solution as bringing real-world assets to the blockchain, creating a "crypto-copy" of the required asset. So, the solution to use bitcoin on the Ethereum blockchain was creating an asset on Ethereum with ERC-20 standard (**E**thereum **R**equest for **C**omments 20) and "pretend" that it is bitcoin. There are several reasons to aim and promote interoperability between the two chains and of blockchains, in general [25]. First, despite the interest of investors in Ethereum and altcoins, the highest portion of liquidity is still retained by bitcoin [26]. Therefore, while in centralized exchanges, almost all the trading pairs are with bitcoin, in decentralized exchanges, the trading pairs are with Eth, Sol, Trx, Waves, and so on, reaching together a much lower volume of trade than bitcoin alone [27]. This creates a lack of liquidity for decentralized exchanges with respect to centralized exchanges. Second, blockchains such as bitcoin are characterized by low scalability and high transaction fees during congestion time. Therefore, it is not suitable for a constant large volume of transactions or microtransactions [28]. Actually is already unprofitable to move an amount lower than 100$ on the bitcoin network. Lastly, crypto belonging to different ecosystems (e.g., Polkadot, Ethereum, Tron) does not adhere to the same standards. Therefore, they are not suitable for being managed by the same smart contract. Furthermore, even in the same ecosystem, tokens may belong to different standards (e.g., ERC-20, ERC-223, TRC-10, TRC-20, BEP2), resulting in the same lack of interoperability.

Bridging separate blockchain and especially bridging bitcoin to other blockchains would solve the problem of liquidity that decentralized finance is experiencing. The capital hold in bitcoin could also be finally used to obtain an interest over its simple appreciation. Bridging blockchains would also decrease latency and the cost of transactions. Bitcoin network has slow block time and high fees (1/10 block/minute, and around 30$/transaction), while using bitcoin on the Tron network, for example, would mean moving them almost for free with a block time of roughly three seconds [29], [30]. In the end, the bitcoin standard only allows basic scripts and not smart contracts, making it impossible to operate on the Bitcoin network the same kind of DeFi contracts available on Ethereum. Although the use of wrapped tokens is still debated in online forums [31], their use is perceived useful to address the above-mentioned issues while an appropriate cross-chain solution is ever developed.

Minting copies of bitcoin (or other crypto assets) on a non-native blockchain such as Ethereum is not a difficult task. The troublesome and difficult part is to ensure the peg of the minted token to the wrapped asset and that this peg is maintained over time. At the moment, three are the means through which the peg is enforced.

Centralized: This method relies on one or more trusted organizations to maintain the wrapped token's value. The third parties are in charge of providing the so-called Proof of Asset (PoA). The PoA proves that the locked assets are safely stored and not used in any other applications. The auditability must be kept over time without any access restriction, and the companies have to ensure that the wrapped assets can be redeemed for the corresponding assets at any time [3], [10].

Trustless: Wrapped tokens can be issued by fully decentralized entities such as smart contracts. They automatically manage the request, minting, burning, and custody of tokens. Being smart contracts, they are naturally trustless and ensures on-chain proof of assets [6], [32].

Hybrid: This method uses both a centralized entity and a smart contract to issue wrapped tokens. The presence of a centralized entity is usually required to perform necessary tasks that smart contracts are still unable to do (e.g., KYC).

Synthetic: This method slightly differs from the previous one as it does not require the lock of the original asset in a trusted vault or a smart contract. In order to mint a synthetic wrapped token, the user needs to lock an amount of assets of the equivalent value or more. Intuitively, when the user wishes to burn the wrapped token, will not receive the represented token back but the provided assets instead [5].

Those are the known methods for issuing wrapped tokens and ensuring the peg to the original asset. However, there are drawbacks to every method that should not be underestimated. Centralized issuance of tokens means that users are fully dependent on the intermediary to keep their assets. Therefore, a considerable degree of trust is required to rely on those institutions and contracts. Trustless alternatives are so-called in the sense that there is not a centralized entity to trust. Users are ensured that no other player can run away with their money or defy the contract. However, as already happened in the past for other smart contracts, those can be bugged or experience malfunction. In that case, although not determined by a central authority, the loss of funds is definitely possible. Hybrid solutions reduce but do not prevent both drawbacks. The presence of a central authority can ensure assistance in case of contract malfunction, and on the other hand, the automation of contracts can ensure a higher level of operation transparency. Again the problem of centralization or bugs still remains. Synthetic wrapped tokens are quite controversial and will be discussed in a dedicated paragraph. The next section analyzes practical cases of wrapped tokens to distinguish the main typologies better.

## 3. Projects of wrapped tokens

### 3.1 Wrapped Ether (WETH)

Although counterintuitive, on the Ethereum network, it also exists a wrapped version of Ethereum itself called WETH. It is actually the first wrapped token ever created and became live in January 2018. Since decentralized platforms running on Ethereum uses smart contracts to facilitate trades, every asset should conform to the same standardized format. This format is recognized as ERC-20. Since Ether was built before the ERC-20 format, it is not suitable for DeFi smart contracts, and for this reason, a tokenized version of ETH (WETH) was perceived necessary [11]. The wrapping of Ether is quite easy and intuitive when compared to other coins because Ether and WETH belong to the same ecosystem. Unlike the wrapping of other assets and cryptocurrencies, no interoperability issues arise since no cross-chain transaction is needed. The need to wrap Ether, in this case, is given by the necessity of changing its standard. As shown in figure 1, the process is a simple swap between Ether and WETH which is made at 1:1 plus gas fees paid in eth. On the other hand, WETH as an ERC-20 token can no longer be used to pay gas fees. The swapped Ether is then held in the smart contract, which is publicly auditable [33]. The process is completely decentralized and, unlike other wrappings, does not involve third-party trusted custodial services. Although WETH, is widely used, as stated on the WETH website, "Hopefully, there is no future for WETH" [11]. There is the intention, in fact, to make Ether compatible with ERC-20 standards. However, other standards such as ERC-223 are also arising.

Finally, It is important to point out that today the acronym WETH is also used for Ether, used on other blockchains. Therefore, WETH standard is actually heterogeneous (e.g., TRC-20, BEP-20).

Figure 1. WETH minting

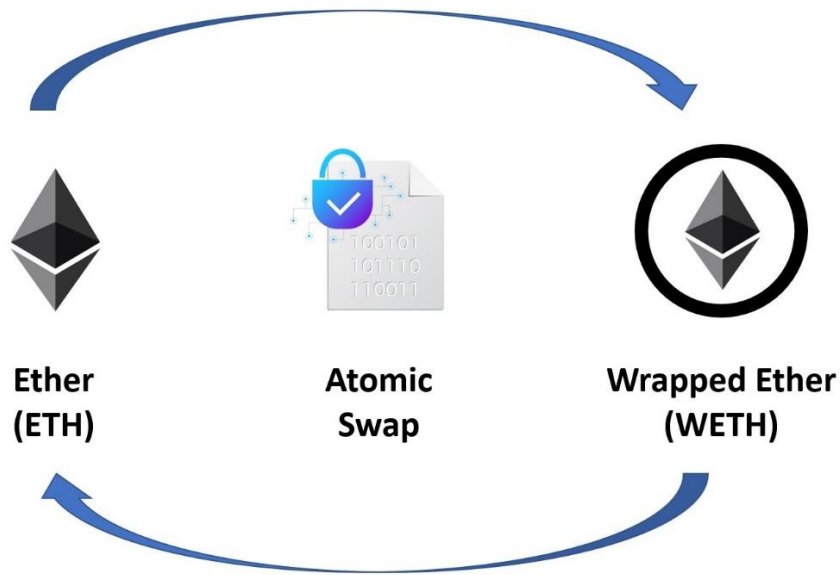

### 3.2 Wrapped Bitcoin (WBTC)

Although WETH was the first wrapped token by chronological order, the concept of wrapped tokens is associated with the announcement in late 2018 of a partnership between Kyber Network, BitGo, and Republic Protocol [34] to launch a wrapped version of bitcoin. Kyber Network is a decentralized liquidity aggregator proposing the best exchange rates from the supported Dexes, such as Skyscanner proposes the best prices for flights scanning all the supported flight companies [35]. BitGo, on the other hand, is a company specialized in custodial services for cryptocurrencies and enabling institutional investments in the crypto space [36]. The Republic Protocol is finally a company specialized in the transaction and swap between bitcoin and token of the Ethereum network. It is focused on the handling of a large quantity of capital offering OTC services [37]. The three companies finally launched the WBTC project in January 2019, when the DeFi sector was still immature but managed to attract 1$ Billion of collateralized assets at his launch and even 7$ Billion by May 2021 [29]. The idea promoted by those companies was to create a stablecoin pegged 1:1 at the value of a crypto asset that could be utilized on a non-native blockchain. The proposed ecosystem to handle WBTC was composed of the following entities:

**Custodian**. In this case, BitGo, has the task of hold and monitor the token delivered as collateral. It is a specialized company that is meant to offer the highest security standards. It has to provide all the necessary transparency to perform auditing procedures on the collateralized assets. The custodians actually perform the "wrapping and unwrapping" part of the whole process [29]. It can either be a centralized company or also a smart contract, such as in the case of WETH [23].

**Merchant**. It is the intermediary between the custodian and the user (figure 2). For WBTC this role was initially managed by Kyber and Republic Protocol. It receives the request from the user and delivers the collateral to the custodian. When the minting process is completed, it also delivers the wrapped token to the users. The merchant is also involved in the reverse process when receives the burning request from the user. In that case, it handles the wrapped token to the custodian returning the unwrapped token to the user [24]. In the case of WBTC, the custodian is also the entity in charge of executing KYC procedures.

**User.** is the entity in need of a wrapped token. It can be a private user or an institution. Given the characteristics of a wrapped token, the scope of the conversion is mainly speculative as the token loses all the original features in exchange for interoperability. The user initiates the request to the merchant that in turn involves the custodian. Basically, to receive the wrapped token, the user has to entrust his tokens to the merchant and pay the fees to both merchant and custodian. In case the user wishes to receive his token back, it has to initiate the burning request and pay the fees again.

**Governance** In the specific case of WBTC the governance of the contract is managed by an "M of N" multi-signature wallet whose key owners are the member of the so-called "WBTC DAO". The governance is in charge of handling the contract and the addition and removal of custodians and merchants. All custodians and merchants will be DAO members, but other institutions can also be included without a custodian or merchant role. In "M of N" wallets, M is the required number of signatures, and N is the total number of members. The values of M and N will be decided mutually between members keeping in mind security as well as the ease of adding/removing members [30].

Figure 2. WBTC minting process

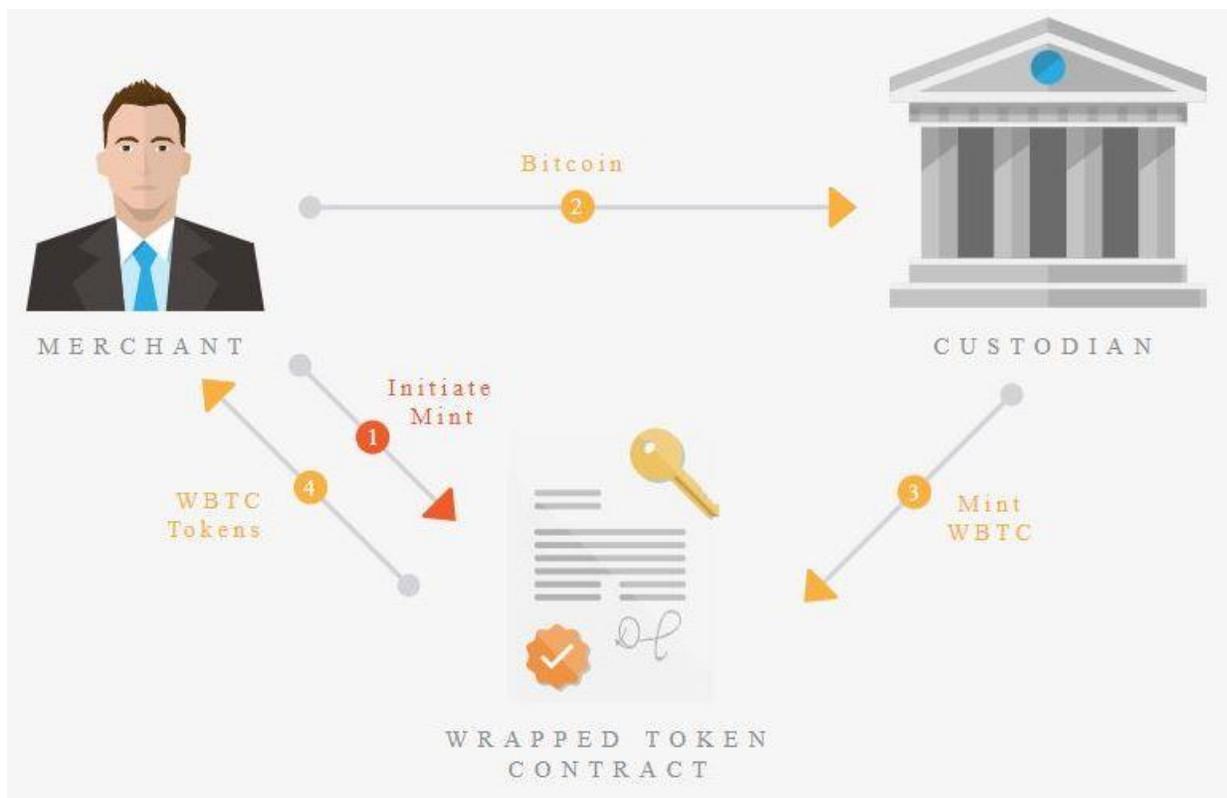
source: https://wbtc.network/

*3.3 Ren Project*

The Republic Protocol (one of the WBTC DAO members) has launched the Ren Project to promote cross-chain interoperability [38]. Whilst Republic Protocol is also involved in the minting process of WBTC, the bitcoin created within the RenVM application are named renBTC. The difference between the minting process of WBTC and renBTC is that it is completely automated, and users can directly bridge their crypto to a different blockchain (e.g., Ethereum, BinanceSmartChain), without performing KYC procedures. Assets are sent to the custodian's address, and the smart contract handles all the processes as well as the fees. The minted wrapped assets are then delivered at the user address corresponding to the destination ecosystem. The RenVM smart contract automates the role of the merchant and the custodian. Moreover, suppose a user wishes to transfer a renBTC from one ecosystem to another(e.g., from Ethereum to Solana). In that case, the protocol just burns the issued tokens and mints the same quantity (-fees) on the newly requested one. At the moment, the project supports few assets and ecosystems, but the developer teams declared the intention to add more [39]. It is important to note that although WBTC and renBTC, are representations of the same token (BTC) which in principle is fungible, the difference in the name with respect to WBTC plays a crucial role. By the time the user has his assets wrapped on the designed network acquires the right to invest them and to trade those for other tokens. However, selling a WBTC does not have the same meaning as selling BTC. What a holder of WBTC transfers by selling the token is the right to redeem the locked token for the same amount of owned wrapped tokens. We recall that the ability to redeem the locked asset is the reason why the wrapped token maintains the exact peg. Therefore since the locked token may be in "custody" of a different entity, it is vital to distinguish those by naming them differently to ensure their non-fungibility. WBTC, for example, cannot be redeemed through the RenVM, since they do not have the corresponding tokens in their custody.

Following the ren Project, other companies such as Binance launched its Project Token Canal to facilitate the issue and binding of more token assets on Binance Chain and Binance Smart Chain, and guarantee the conversion from and to the original tokens. The Binance bridge, for example, works very similarly to the RenVM, pointing, however, to different ecosystems (BinanceSmartChain, Ethereum, Tron) [20]. Above their decentralized bridge, however, users can receive wrapped tokens directly from the Binance exchange, withdrawing crypto to a non-native platform supported by Binance. However, there are two consequences to keep in mind when using this feature from a centralized exchange. Being KYC mandatory on centralized exchanges, wrapped tokens will not be acquired privately. Second, if wrapped tokens are obtained from an exchange, the existence of a corresponding locked asset is also not guaranteed. However, as declared, by binance, all the funds and issued tokens are covered and insured by the so-called SAFU fund, which should correspond to 10% of all the deposits [40].

### 3.4 Synthetic Tokens and sBTC

A synthetic token is conceptually different from a wrapped token, although it may have the same standard (ERC-20) and probably the same price. The synthetic Bitcoin, is, for example, a derivative of the original token; therefore, its only function is to speculate on its price. We recall that a derivative is an asset that derives its value from an underlying asset or index [41]. Unlike the wrapped token, a synthetic token cannot be redeemed for the original token since no original token was locked in order to mint the synthetic version. However, the mechanism to issue a synthetic token such as sBTC is similar to the wrapped token. On the Synthetix platform, for example (figure 3), in order to mint sBTC, 800% of the minted asset has to be deposited in SNX token as collateral [4]. An oracle service is queried in order to obtain the BTC/USD exchange rate, which in the case of Synthetix is Chainlink. The collateral is then returned when the user returns the sBTC, which is consequently burnt. It is also possible to acquire a short position on sBTC depositing collateral in sUSD. In that case, however, the fluctuation on the sBTC price will affect the collateral deposited depending on the BTC price. Therefore, the quantity of sUSD redeemed will differ from the provided amount (-fees). The advantage of utilizing a synthetic token is that the process is completely trustless and automated. Unlike the minting process of WBTC and renBTC, the issuance of sBTC does not involve a custodian that locks the crypto on a different chain. The process is automated by the smart contract and involves already interoperable assets. In this sense, the issuance of sBTC is more similar to the issuance of WETH. However, the fact that the original token is not locked to mint the synthetic assets makes those tokens not as accepted as the wrapped versions [14]. It must be said that the synthetic token in its structure is not different from the wrapped token and actually has the same functionalities. What differs is its source and its acceptance.

Figure 3. Synthetic BTC minting

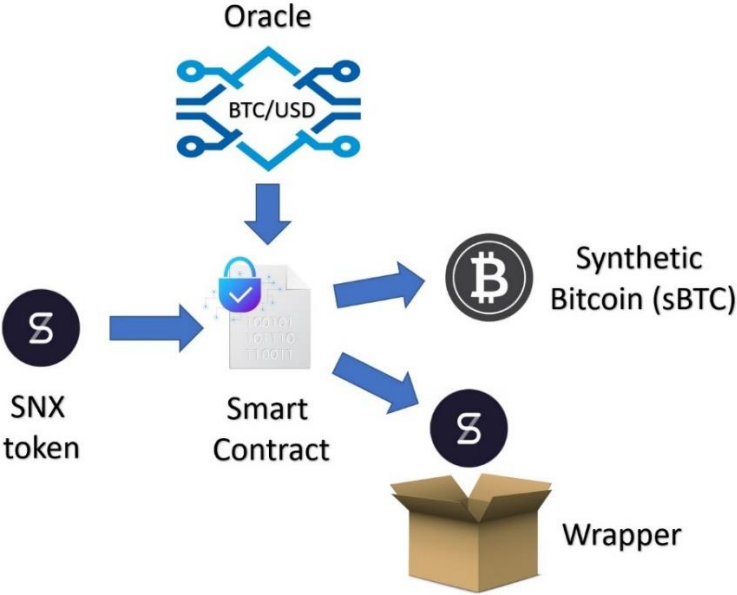

*3.5 Secret Network*

Since the creation of bitcoin, the cryptocurrency world has had the task of balancing privacy and transparency. As blockchains are mostly public, we are witnessing the rise of firms specializing in auditing blockchain transactions. Their aim is to discover the owner of addresses and the motives of transactions. Due to that increasing lack of privacy, in 2014, the Enigma project was founded with the aim of providing privacy to smart contracts on Ethereum. In 2020 the enigma launched its own main-net with the name of Secret Network [42]. The secret network is based on Cosmos SDK, so it has the same consensus mechanism (delegated proof of stake) and the same block-time (6-7 seconds). Similarly, to the Binance chain, it consists of two parallel ecosystems, of which one is public, and one is private. The public ecosystem is mainly used to transfer and stake SCRT native tokens. Instead, the private ecosystem is used for privacy-preserving applications developed on their protocol and to transact secret tokens. The most innovative part of the secret network is the implementation of a bridge that is very similar to the RenVM to enable interoperability between different ecosystems [43]. Instead of just wrapping tokens to be usable in another ecosystem, the secret network empowers a dedicated layer two ecosystem where all the wrapped tokens can interact (figure 4). Unlike RenVM, the secret bridge directly connects wallets supporting separate ecosystems (e.g., Ethereum and Secret). The users select the token and quantity that he wishes to bridge, and once the process is done, a snip-20 version (native token standard) of the locked token is minted on the secret network. The original token is locked in a smart contract on its platform, and the bridge should ensure its safety and auditability. Although the various tokens can interact together and be inserted in liquidity pools, those, just like WBTC and renBTC are non-fungible. It means that although with the same standard and on the same platform, tokens coming, for example, from the Ethereum platform cannot be redeemed on the Binance Chain. Therefore, a holder with two equal tokens coming from different ecosystems cannot merge them in his secret wallet. As explained for the Ren Project, this feature is crucial to ensure that the wrapped tokens are redeemed for the corresponding locked asset [15].

Figure 4. Interoperable layer two network

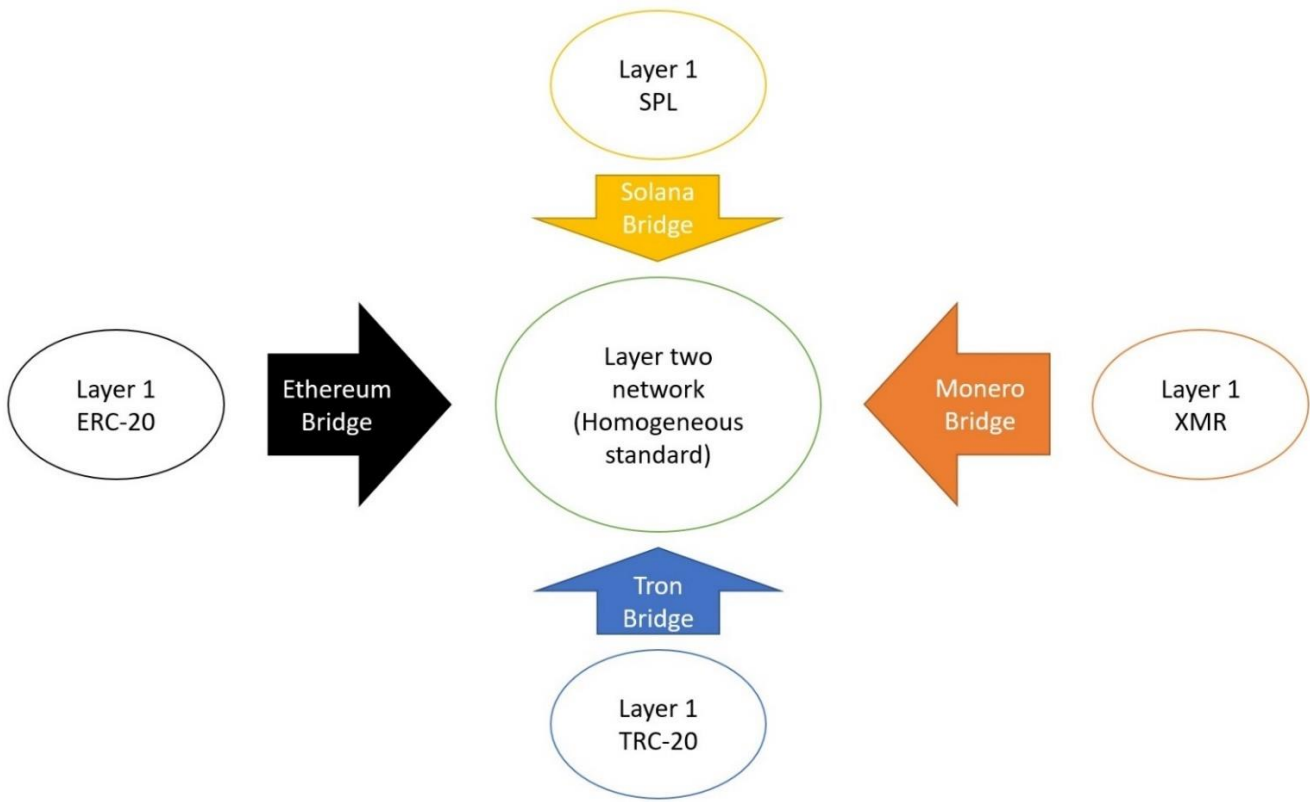

## 4. Discussion

The need for wrapped tokens is justified by the inability of blockchains to directly communicate with each other to overcome the lack of interoperability [1]. As described, the mean of minting wrapped tokens is very similar to that of minting stablecoins. Unlike custodial stablecoins, however, the minting is initiated by the users and not by institutions. On the other hand, they cant be considered similar to non-custodial stablecoins due to the centralized nature of custodians [17]. The utilization of custodians is a very debated concept in the crypto space. As a matter of fact, the whole idea of decentralization and DeFi, should involve the ability of users to have complete control of their assets without involving trusted third parties and without KYC requirements [44]. Due to AML reasons, custodians of crypto assets are obliged to request KYC from users depositing collaterals. Therefore, the process of minting most of the wrapped tokens cannot be initiated without a KYC. As centralized entities, custodians activity is subject to regulation and may also be ceased or censored [10], [30], [45].

Users also have the choice of acquiring wrapped tokens privately on a decentralized exchange without submitting KYC procedures. However, as explained, the intrinsic value of the wrapped token lies in the right to redeem the corresponding locked asset. Therefore, although the token may have been acquired privately, KYC procedures cannot be avoided at the moment of token redemption. If in order to redeem WBTC the user has to undertake KYC procedure, the benefits of privacy given by the blockchain technology are forfeited. Projects such as Binance Bridge and RenVM, instead, do not require KYC when redeeming tokens [20]. However, as RenVM terms and conditions stated, regulations may always change procedures or even delete some [45]. This condition means that although the contract is apparently trustless and decentralized, the company managing the protocol may be ceased or be obliged to change some function at any time due to regulatory hurdles. Among those discussed in the present article, the only wrapped tokens characterized by privacy and trustless minting process appear to be synthetic tokens and those issued on Secret Network [14], [43].

As already explained, the issuance of synthetic tokens is very similar to wrapped tokens. However, they are often non considered as such. Wrapped tokens ensure that the circulation of a specific token remains the same regardless of the chain on which it is operating. It means that if a WBTC is being lended on Ethereum, a BTC is contextually being frozen on the Bitcoin blockchain. This ensures that the supply remains the same and the price of both assets is safeguarded. For synthetic assets, however, this condition does not hold. As an example, synthetic bitcoin operating on Ethereum, does not have a counterparty locked on the Bitcoin network, de facto altering the total supply of BTC. This may be a reason why wrapped tokens still have a wider acceptance in DeFi platforms such as Compound and MakerDAO, with respect to synthetic tokens. Another important aspect to consider, which is rarely discussed, is the smart contract reliance on external oracle services [46]. Since the locked token is not the same as the synthetic token, an oracle service has to be queried to obtain the token's exchange rates. As discussed in many papers, those oracle services are centralized entities prone to malfunction and manipulation [47], [48]. Therefore, even if the synthetic contract is trustless and automated, the reliance on a third-party service constitutes a single point of failure.

Finally, the approach implemented by Secret Network is lately being seen in other ecosystems such as Tron and BitTorrent [8]. Their idea is basically to create a layer two environment in which all the tokens have the same standard and can interact together. If this approach enables interoperability, on the one hand, it has considerable limitations on the other. First, as a stand-alone environment, it is not compatible with all the most prominent applications in the DeFi space based on Ethereum. The company is, in fact, developing its own applications, which at the moment counts less than 0.07% of the total capital invested in DeFi [7]. Convincing investors to switch from their trusted platform to a newly built one may not be an easy task. Second, the availability of wrapped assets is subject to the development of bridges. Bitcoin bridge, for example, is still not available; therefore, it is not possible to directly interact with bitcoin on the secret network. However, it is possible to bridge WBTC and renBTC from the Ethereum network. This option would mean wrapping the original token (bitcoin) actually two times. Third, as stated in the secret network F.A.Q., using a layer two network is not free of risks. Being a smaller network compared to Ethereum and bitcoin is more subject to attacks or unavailability of nodes. In case of network failures, there is the chance to lose all the tokens locked in the bridge [49].

In the end, it is arguable that unlike bitcoin or dash, which can be considered decentralized currencies, a wrapped token is mainly issued and managed by organizations that "guarantee" their value and reconvertibility to the former token. The trust to put in the token issuer and the token itself is strictly related to the specific mechanism undertaken to mint the token, which varies according to the issuing company/authority.

Table 1. Approaches to interoperability: advantages and drawbacks

| Solution | Example | Advantages | Drawbacks |
| --- | --- | --- | --- |
| Centralized | WBTC | -Non-native blockchain support<br>-Smart contract support<br>-High scalability | -Native blockchain support lost<br>-Different security standards<br>-Different fees mechanisms<br>-Loss of Privacy<br>-Loss of decentralization<br>-Chance to lose the locked token<br>-high minting fees |
| Decentralized | WETH | -compliance with non-native standard<br>-Interoperability with smart contracts<br>-Automated procedure<br>-Low minting costs | -Loss of native characteristics<br>-Subject to standard obsolescence |
| Hybrid | sWBTC | -Low Minting Costs<br>-Low transaction costs<br>-High scalability<br>-Privacy safe | -Native standards unsupported<br>-Reliance on dedicated applications<br>-Different security standards<br>-Centralized custodians<br>-Chance of losing the locked asset |
| Synthetic | sBTC | -Non-native blockchain support<br>-Heterogeneous collateral supported<br>-Trustless application<br>-Derivative applications supported | -Alter the represented asset supply<br>-Not fully recognized as wrapped token<br>-Unaccepted in many applications or wallets<br>-High minting fees<br>-High collateralization ratio<br>-reliance on external oracle service |

## 5. Conclusion

The scope of this paper was to undertake a study on wrapped tokens. Wrapped tokens are a practical solution for cross-chain communication and interoperability [50]. They also enable faster and more accessible alternatives for transactions when a particular blockchain is experiencing delays or high fees. Allowing the use of tokens over a non-native blockchain can constitute an effective solution to bring liquidity for developing DeFi space [4]. Reducing transaction costs and speed can enable the microtransactions that were initially doable at the birth of blockchain and became soon unmanageable due to congestions and transaction fees. However, there are drawbacks to consider when interacting with those tokens that mainly concern the entities involved in their creation. Custodians, whose role is essential to keep the tokens locked, are centralized entities subject to audits and regulations. If, on the one hand, this guarantees their trustworthiness, on the other hand, the pressure of regulation may alter the proposed service or even censor transactions. Other wrapped tokens may work without a custodian. Its role is de facto, replaced by a smart contract or virtual machine. Their use, although decentralized, is not risk-free since smart contracts are not exempt from errors and bugs [51].

Although the promise of wrapped tokens is to reduce fees, the truth is that this is not guaranteed. Wrapping bitcoin to Ethereum would result first in high minting costs, and then transactions would be subject to Ethereum fees that are not always lower than the bitcoin one [10]. Therefore, switching chain every time to reduce transaction costs seems not a convenient option. Nevertheless, a wrapped version of a coin is not that coin. It's another cryptocurrency entirely. Therefore, most of their functionalities are forfeited. For example, some passive income platforms don't take wrapped tokens, and neither do some exchanges and wallets. As explained, although counterintuitive, it is not possible to pay gas fees with WETH on the Ethereum network even if it is a wrapped token of Ether [50].

As for the currently available technology, wrapped tokens can't be used for true cross-chain transactions [32]. Just as in the case of oracles, the first approach to the problem is to reintroduce centralization in order to overcome the encountered issues. The examples discussed in this paper show that the offered proposals constitute a mere workaround to the problem and not a definitive solution. In every case, there are some features to give up in exchange for interoperability. While WBTC needs trust over custodians, renBTC is eventually subject to smart contracts and regulatory pressures. Synthetic solutions are often debated for their minting process and their impact on the market. On the other hand, layer two solutions such as Necret Network require trust in a different network with different security and applications.

The shared opinion is that the use of the wrapped token is temporary, and it will probably be abandoned when a true cross-chain solution is discovered [52]. Since it is not sure when and if full interoperability between blockchain may be reached, further studies over the security of interoperability tokens could be useful to understand their development and potential better.

**References.**